\documentclass[12pt,preprint,showpacs,preprintnumbers]{revtex4}
\usepackage{amssymb}
\usepackage{amsmath}
\usepackage{graphicx}
\usepackage{subfigure}
\usepackage{dcolumn}
\usepackage{bm}
\usepackage{epsfig}
\usepackage[T1]{fontenc}
\usepackage{ae,aecompl}

\begin{document}

\title{Cellular automata model for elastic solid material\footnote{This work is supported by the Science Foundations of CAEP
under Grant Nos. 2012B0101014 and 2011A0201002, National Natural
Science Foundation of China under Grant Nos. 11075021, 91130020 and 11202003.}}
\author{Yinfeng Dong$^{1}$, Guangcai Zhang$^{1}$\footnote{%
Corresponding author. Email address: Zhang\_Guangcai@iapcm.ac.cn},
Aiguo Xu$^{1}$\footnote{%
Corresponding author. Email address: Xu\_Aiguo@iapcm.ac.cn}, Yanbiao Gan$^{2}$}

\affiliation{
1, National Key Laboratory of Computational Physics, \\
Institute of Applied Physics and Computational Mathematics, P. O.
Box 8009-26, Beijing 100088, P.R.China\\
2, North China Institute of Aerospace Engineering, Langfang 065000,
P.R.China }

\date{\today }

\begin{abstract}
The Cellular Automaton (CA) modeling and simulation of solid dynamics is a long-standing difficult problem.
In this paper we present a new two-dimensional CA model for solid dynamics. In this model the solid body is represented by a set of white and black particles alternatively positioned in the $x$- and $y$- directions. The force acting on each particle is represented by the linear summation of relative displacements of the nearest-neighboring particles.  The key technique in this new model is the construction of eight coefficient matrices.
Theoretical and numerical analyses show that the present model can be mathematically described by a conservative system. So, it works for elastic material. In the continuum limit the CA model recovers the well-known Navier equations. The coefficient matrices are related to the shear module and Poisson ratio of the material body.
Compared with previous CA model for solid body, this model realizes the natural coupling of deformations in the $x$- and $y$- directions. Consequently, the wave phenomena related to the Poisson ratio effects are successfully recovered. This work advances significantly the CA modeling and simulation in the field of computational solid dynamics.
\end{abstract}

\maketitle

\section{Introduction}

Cellular Automaton (CA) is a kind of discrete model for complex systems containing large numbers of simple components
with local interactions \cite{FHP,Wolfram_Physica10D,Langton-AL,Wolfram-Book,Chopard-book}.
It is charming that CA is able to present macroscopic complex
phenomena with simple rules. The interesting fact hints that CA
can provide a new conceptual framework, as well as an effective
numerical tool, which can be used to explore the microscopic or mesoscopic mechanism of
physical systems by simulating macroscopic behaviors. Further, we
can capture more essential features of given physical systems by
amending the rules. Therefore, the CA approach and the related modeling techniques are
powerful methods to describe, simulate and understand physical
systems. During the past decades CA models have been
introduced and studied in various fields \cite{FHP,Wolfram_Physica10D,Langton-AL,traffic-flow,Quantum_dot-Sci,snow-avalanche,Succi,Wolfram-Book,Chopard-book,Jiaoyang},
such as the artificial life, traffic flow, quantum computation, snow avalanche, disease control and prevention, etc.

Historically, Frisch \emph{et al.} \cite{FHP} obtained the
correct Navier-Stokes equations starting from the lattice gas CA and
introduced CA approach into the field of fluid dynamics in 1980s, since then
the CA approach has been particularly successful in this field. It
is meaningful to mention that study of lattice gas CA has also inspired
the appearance of the well-known Lattice Boltzmann (LB) method which has
attracted extensive attention and has been successfully applied in
various fields \cite{Succi,Xu-PRE,HaipingFang}.

Besides the fields mentioned above, some researchers have developed the CA/LB approach to study
viscoelastic fluid and solid. For example, Giraud, \emph{et al.} proposed an LB model for viscoelastic fluid \cite{Viscolastic fluid}. Mora,
\emph{et al.} presented their LB phononic lattice solid
in Ref. \cite{Phononic LB}. Fang, \emph{et al.} proposed an LB model for photonic band gap
materials \cite{Photonic band gap}. Xiao studied the shock wave
propagation in solids by LB method \cite{Xiao}. Chopard and Marconi \emph{et al.} proposed a CA
of solid which can be thought of as a grid of masses linked by
springs \cite{Chopard-book,Marconi PhD thesis}. Based on the CA,
they developed a kind of LB method for solids. Matic and  Geltmacher developed a CA for modeling the mesoscale damage evolution \cite{fracture}.

However, in the field of solid mechanics, due to some unsolved problems, the CA approach has not been as successful as in the field of fluid dynamics. For instance,
Chopard, \emph{et al.} presented a very nice idea to recover the elastic behavior, but their CA still can not couple the motion in $x$ direction with that in $y$ direction.
This limitation furthers to plague describing the Poisson ratio of solid materials and more
practical behaviors. In order to simulate the dynamics of elastic solid material, it is necessary to modify the solid CA of Chopard, \emph{et al.} to couple
the motions in different directions. This is the main aim of the present paper.

The following parts of the paper are planned as follows. In the next section, we briefly review Chopard's CA, describe the improved model and perform numerical
stability analysis for the new model. Some numerical examples are given in Sec. III to validate the model.
Conclusions are summarized in Sec. IV.

\section{Cellular Automata model for Elastomer}

\subsection{Chopard's Cellular Automata}

A sketch of the two-dimensional particle set for Chopard's CA is shown in Fig. 1. In the case of translation motion,
the automata rules are: (i) Divide the lattice into two
sublattices (black/white) in a checkboard manner; (ii) Alternatively
update all particle positions in each sublattice. Considering only the nearest neighboring inter-particle interactions gives the following rules:
\begin{equation}
\left\{
\begin{aligned}
{\bf b}^+=-{\bf b}+\frac{2}{K} \sum_i {\bf w}_i,\\
{\bf w}^+=-{\bf w}+\frac{2}{K} \sum_i {\bf b}_i^+,  \label{eq1}
\end{aligned}
\right.
\end{equation}
where $\bf{b}$ and $\bf{w}$ are the displacements of black particles
and white particles relative to their initial equilibrium positions, respectively; $\textbf{w}_i$ are the displacements
of black particles's neighbor and vice versa; $K$ is the number of
a cell's neighbor; ``+" denotes the next time step. In the case of
rotation motion, the automata is quite difficult to describe the two-dimensional and three-dimensional
behaviors. Through Chopard's CA above, we can see some mesoscopic mechanisms
working in the framework of cellular automata modeling. In this section, we aim to obtain
a CA that can represent the behaviors of elastomer ruled by Navier equation
\begin{equation}
\partial ^2_t \bf{u}=(\lambda+\mu)\nabla(\nabla \cdot \bf{u})+\mu
\nabla^2 \bf{u}.
\end{equation}

\begin{figure}
\includegraphics[width=0.4\textwidth]{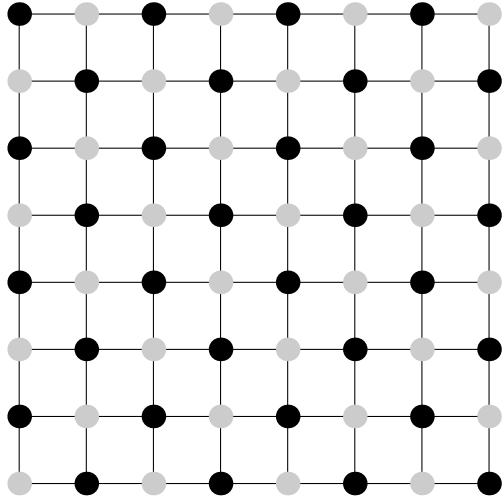}
\caption{Grids for the two dimensional CA.}
\end{figure}
\begin{figure}
    \includegraphics[width=0.50\textwidth]{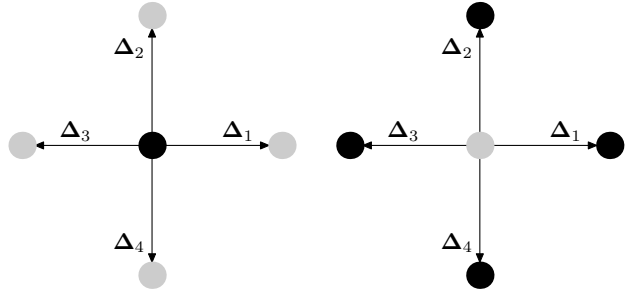}
   \caption{Directions of displacement vectors of a particle.}
\end{figure}

\subsection{Our new Cellular Automata for elastomer}

In the present work, we focus on the two dimension CA model, which is capable of coupling the motion
in $x$-direction with that in $y$-direction and capable of representing the Poisson ratio of solid materials.
We use the same particle set as shown in Fig. 1.
The rules of CA are as follows: (i) Every particle updates its position according to the
neighbors's current states; (ii) Subgrids (black and white)
alternatively update particles' positions. Obviously, every
particle has four neighbor particles in this model.
The directions of every cell's neighbors are shown in Fig. 2,
where ${\Delta}_i$ ($i=1,2,3,4$) is the neighboring vector at the initial time. Now we can modify Eq. (1) to the following form
\begin{equation}
\left\{
\begin{aligned}
{\bf b}^+&=-{\bf b}+\sum_i {\bf B}_i {\bf w}_i,\\
{\bf w}^+&=-{\bf w}+\sum_i {\bf D}_i {\bf b}_i^+,
\end{aligned}
\right.
\end{equation}
where ${\bf B}_i$ are the coefficient matrices of black particles' neighbors,
${\bf D}_i$ are the coefficient matrices of white particles'
neighbors. It is noteworthy that Eq. (3) will degenerate into
Eq. (1) when ${\bf B}_i={\bf D}_i=I/2$.

From the basic evolution Eq. \eqref{eq1}, we can obtain the following
volume ratio of consecutive two evolution steps in the phase space
\begin{equation*}
\frac{\partial \left( \left\{ \mathbf{b}^{+}\right\} ,\left\{ \mathbf{w}%
^{+}\right\} \right) }{\partial \left( \left\{ \mathbf{b}\right\} ,\left\{
\mathbf{w}\right\} \right) }=1\text{,}
\end{equation*}%
where $\left\{ \mathbf{b}\right\} =\left\{ \mathbf{b}_{i}\text{, }i=1\text{,}%
2\text{,}\cdots \right\} $ represents displacements of all black particles, $%
\left\{ \mathbf{w}\right\} =\left\{ \mathbf{w}_{i}\text{, }i=1\text{,}2\text{%
,}\cdots \right\} $ represents displacements of all white particles. It is
clear that the CA system is conservative.

Let $S_i=e^{\lambda {\bf \Delta}_i \cdot \nabla}$, Eq. (3) can be
written as follows
\begin{equation}
\left\{
\begin{aligned}
{\bf{b}}^+=&-{\bf b}+\sum_i {\bf B}_i S_i \bf{w},\\
{\bf{w}}^+=&-{\bf w}+\sum_i {\bf D}_i S_i \bf{b}^+.
\end{aligned}
\right.
\end{equation}
Furthermore, if we let $ T=e^{\lambda \delta \partial_t}$, Eq. (4) can be rewritten
as Eq. (5)
\begin{equation}
\left\{
\begin{aligned}
(T+1){\bf{b}}=& \sum_i {\bf B}_i S_i \bf{w},\\
(T+1){\bf{w}}=& \sum_i {\bf D}_i S_i T\bf{b}.
\end{aligned}
\right.
\end{equation}
To remove the quantity ${\bf w}$ by combining equations, we
obtain
\begin{equation}
(T+1)^2 {\bf b}= \sum_{i,j} {\bf B}_i S_i {\bf D}_j S_j T {\bf b}.
\end{equation}

To perform Taylor expansion with $\lambda$ on the both sides of Eq.
(6) and compare the same order terms, we can obtain a series of
equations.

The zero-order term
\begin{equation}
{\bf BD}=4{\bf I},
\end{equation}
where ${\bf B}=\sum_i {\bf B}_i$ and ${\bf D}=\sum_i {\bf D}_i$.

The 1st-order term
\begin{equation}
2(T+1)T^{\prime}{\bf b}= \sum_{i,j} ({\bf B}_i S_i^{\prime}{\bf D}_j S_jT+{\bf
B}_i S_i{\bf D}_j S_j^{\prime}T+{\bf B}_i S_i{\bf D}_j S_j
T^{\prime}){\bf b}.
\end{equation}
Eq. (8) can be rewritten as
\begin{equation}
4\delta \partial_t {\bf b}=(\hat{\bf B}{\bf D}+{\bf B}\hat{\bf
D}+{\bf BD}\delta \partial _t){\bf b},
\end{equation}
where
\begin{equation}\left\{
\begin{aligned}
\hat{\bf B}=\sum _ i {\bf B}_i {\bf \Delta}_i \cdot \nabla,\\
\hat{\bf D}=\sum _ i {\bf D}_i {\bf \Delta}_i \cdot \nabla.
\end{aligned}\right .
\end{equation}
Combining Eqs. (7) and (9), we have
\begin{equation}
\hat{\bf B}{\bf D}+{\bf B}\hat{\bf D}=0.
\end{equation}
Expanding Eq. (11), we obtain
\begin{equation}
[({\bf B}_1-{\bf B}_3){\bf D}+{\bf B}({\bf D}_1+{\bf
D}_3)]\partial_x+({\bf B}_2-{\bf B}_4){\bf D}+{\bf B}({\bf D}_2+{\bf
D}_4)]\partial_y=0.
\end{equation}

The 2nd-order term
\begin{equation}
\begin{aligned}
2(T+1)T''{\bf b}+2T'^2{\bf b}&= \sum_{i,j} ({\bf B}_i S_i''{\bf D}_j S_j T+{\bf
B}_i S_i{\bf D}_j S_j'' T+{\bf B}_i S_i{\bf D}_j S_j T''){\bf
b}\\
&+2 \sum_{i,j} ({\bf B}_i S_i'{\bf D}_j S_j' T+{\bf B}_i S_i'{\bf D}_j S_j
T'+{\bf B}_i S_i{\bf D}_j S_j' T'){\bf b}.
\end{aligned}
\end{equation}
Eq. (13) can be rewritten as
\begin{equation}
4(\delta \partial_t)^2{\bf b}+2(\delta \partial_t)^2{\bf
b}=[\hat{\hat{{\bf B}}}{\bf D}+{\bf B}\hat{\hat{{\bf D}}}+{\bf
BD}(\delta \partial_t)^2]{\bf b}+2(\hat{\bf B} \hat{\bf D}+\hat{\bf
B}{\bf D}\delta \partial_t+{\bf B}\hat{\bf D}\delta \partial_t){\bf
b},
\end{equation}
where
\begin{equation}\left\{
\begin{aligned}
\hat{\hat{{\bf B}}}=\sum_i {\bf B}_i({\bf \Delta}_i \cdot \nabla)^2,\\
\hat{\hat{{\bf D}}}=\sum_i {\bf D}_i({\bf \Delta}_i \cdot \nabla)^2.
\end{aligned} \right.
\end{equation}
Eq. (14) can be changed into
\begin{equation}
2(\delta \partial_t)^2{\bf b}=(\hat{\hat{{\bf B}}}{\bf D}+{\bf
B}\hat{\hat{{\bf D}}}){\bf b}+2\hat{\bf B}\hat{\bf D}{\bf b}.
\end{equation}
Expanding Eq. (16), we have
\begin{equation}
\begin{aligned}
\partial ^2_t {\bf b} = \frac{\Delta ^2}{2\delta^2}&\{\partial^2_x[({\bf B}_1+{\bf B}_3){\bf D}+{\bf B}({\bf D}_1+{\bf D}_3)
+2({\bf B}_1-{\bf B}_3)({\bf D}_1-{\bf
D}_3)]\\
&+\partial^2_y[({\bf B}_2+{\bf B}_4){\bf D}+{\bf B}({\bf D}_2+{\bf
D}_4)+2({\bf B}_2-{\bf B}_4)({\bf D}_2-{\bf D}_4)]\\
&+\partial_x
\partial_y[(({\bf B}_1-{\bf B}_3)({\bf D}_2-{\bf
D}_4)+({\bf B}_2-{\bf B}_4)({\bf D}_1-{\bf D}_3)]\}.
\end{aligned}
\end{equation}

In order to study the deformation of an elastic body, we select the
Navier equation
\begin{equation}
\begin{aligned}
\partial^2_t {\bf b}&=c^2_1 \nabla^2 {\bf b}+c^2_2\nabla \nabla \cdot
{\bf b}\\
&=\partial ^2_x\left(
  \begin{array}{ll}
    c_1^2+c_2^2 &  \\
     & c^2_1 \\
  \end{array}
\right)+\partial ^2_y\left(
  \begin{array}{ll}
    c_1^2 &  \\
     & c^2_1 +c_2^2\\
  \end{array}
\right)+\partial_x \partial_y\left(
  \begin{array}{ll}
     & c_2^2 \\
   c_2^2  & \\
  \end{array}
\right),
\end{aligned}
\end{equation}
where $c_1^2=\mu$, $c_2^2=\lambda+\mu$, $\lambda$ and $\mu$ are Lame coefficients. Physically,
$c_1$ is the sound speed of transverse wave, $\sqrt{c_1^2 + c_2^2 }$ is the sound speed of longitudinal wave.

Combining Eqs. (7), (12), (17) and (18), we get
\begin{equation}
\left\{
\begin{aligned}
 &{\bf D}={\bf B}=2\textbf{I},\\
 &{\bf B}_1-{\bf B}_3+{\bf D}_1-{\bf D}_3=0,\\
 &{\bf B}_2-{\bf B}_4+{\bf D}_2-{\bf D}_4=0,\\
 &\frac{\Delta ^2}{\delta ^2}[{\bf B}_1+{\bf B}_3+{\bf D}_1+{\bf
 D}_3+({\bf B}_1-{\bf B}_3)({\bf D}_1-{\bf D}_3)]=\left(
  \begin{array}{ll}
    c_1^2+c_2^2 &  \\
     & c^2_1 \\
  \end{array}
\right),\\
 &\frac{\Delta ^2}{\delta ^2}[{\bf B}_2+{\bf B}_4+{\bf D}_2+{\bf
 D}_4+({\bf B}_2-{\bf B}_4)({\bf D}_2-{\bf D}_4)]=\left(
  \begin{array}{ll}
    c_1^2 &  \\
     & c^2_1+c_2^2 \\
  \end{array}
\right),\\
&\frac{\Delta ^2}{\delta ^2}[({\bf B}_1-{\bf B}_3)({\bf D}_2-{\bf
D}_4)+({\bf B}_2-{\bf B}_4)({\bf D}_1-{\bf D}_3)]=\left(
  \begin{array}{ll}
     & c_2^2 \\
     c_2^2 &  \\
  \end{array}
\right).
\end{aligned} \right.
\end{equation}
For the sake of convenience, we let $\frac{\Delta ^2}{\delta ^2}=1$,
thus $c_1^2$ and $c_2^2$ are dimensionless. Consequently, Eq. (19) is simplified to
\begin{equation}
\left\{
\begin{aligned}
 &{\bf D}={\bf B}=2\textbf{I},\\
 &{\bf B}_1-{\bf B}_3+{\bf D}_1-{\bf D}_3=0,\\
 &{\bf B}_2-{\bf B}_4+{\bf D}_2-{\bf D}_4=0,\\
 &{\bf B}_1+{\bf B}_3+{\bf D}_1+{\bf
 D}_3+({\bf B}_1-{\bf B}_3)({\bf D}_1-{\bf D}_3)=\left(
  \begin{array}{ll}
    c_1^2+c_2^2 &  \\
     & c^2_1 \\
  \end{array}
\right),\\
 &{\bf B}_2+{\bf B}_4+{\bf D}_2+{\bf
 D}_4+({\bf B}_2-{\bf B}_4)({\bf D}_2-{\bf D}_4)=\left(
  \begin{array}{ll}
    c_1^2 &  \\
     & c^2_1+c_2^2 \\
  \end{array}
\right),\\
&({\bf B}_1-{\bf B}_3)({\bf D}_2-{\bf D}_4)+({\bf B}_2-{\bf
B}_4)({\bf D}_1-{\bf D}_3)=\left(
  \begin{array}{ll}
     & c_2^2 \\
     c_2^2 &  \\
  \end{array}
\right).
\end{aligned} \right.
\end{equation}
Now, we obtain the discrete model described by Eq. (20). By selecting a
series of appropriate matrices $B_i$ and $D_i$, we will obtain a CA
model for elastic body.

To seek such matrices, we introduce another two matrices $\textbf{P}$ and $\textbf{Q}$
that are defined as in Eq. (21)
\begin{equation}
\left \{
\begin{aligned}
{\bf B}_1-{\bf B}_3={\bf D}_3-{\bf D}_1={\bf P},\\
{\bf B}_2-{\bf B}_4={\bf D}_4-{\bf D}_2={\bf Q}.
\end{aligned} \right.
\end{equation}
Thus, Eq. (20) can be rewritten as
\begin{equation}
\left\{
\begin{aligned}
 &4\textbf{I}-{\bf P}^2-{\bf Q}^2=\left(
  \begin{array}{ll}
    2c_1^2+c_2^2 &  \\
     & 2c^2_1+c_2^2 \\
  \end{array}
\right),\\
 &-{\bf P}{\bf Q}-{\bf Q}{\bf P}=\left(
  \begin{array}{ll}
     & c_2^2 \\
    c_2^2 &  \\
  \end{array}
\right),
\end{aligned} \right.
\end{equation}
and Eq. (22) can be changed into
\begin{equation}
4\textbf{I}-({\bf P}+{\bf Q})^2=\left(
  \begin{array}{ll}
     2c_1^2+c_2^2 &  \\
   & 2c_1^2+c_2^2 \\
  \end{array}
\right)+\left(
  \begin{array}{ll}
     & c_2^2 \\
    c_2^2 &  \\
  \end{array}
\right).
\end{equation}
Here we select the forms of ${\bf P}$ and ${\bf Q}$ as follows
\begin{equation}
\left\{
\begin{aligned}
{\bf P}=p_0 {\bf I}+p_1\cdot {\mathbf \sigma}_1,\\
{\bf Q}=q_0 {\bf I}+q_1\cdot {\mathbf \sigma}_1,
\end{aligned} \right.
\end{equation}
where
\begin{equation}
{\mathbf \sigma}_1 =\left(
  \begin{array}{ll}
    0 & 1\\
    1 & 0\\
  \end{array}
\right),
\end{equation}
is the first Pauli matrix.
So we have
\begin{equation}
\left\{
\begin{aligned}
{\bf P}^2+{\bf Q}^2&=(p_0^2+q_0^2+p_1^2+q_1^2){\bf I}+2(p_0 p_1+q_0 q_1) \mathbf{\sigma}_1,\\
{\bf P}{\bf Q}+{\bf Q}{\bf P}&=(2p_0 q_0+2 p_1 q_1){\bf I}+2(p_0
q_1+q_0 p_1){\bf \sigma}_1.
\end{aligned} \right.
\end{equation}
Therefore, Eq. (23) can be changed into algebraic equations
\begin{equation}
\left\{
\begin{aligned}
&p_0^2+q_0^2+p_1^2+q_1^2=4-2c_1^2-c_2^2,\\
&p_0p_1+q_0q_1=0,\\
&p_0q_0+p_1q_1=0,\\
&-2(p_0q_1+q_0p_1)=c_2^2.
\end{aligned} \right.
\end{equation}
After choosing appropriate $c_1$ and $c_2$, we can obtain the
values of $p_0$, $q_0$, $p_1$, and $q_1$, namely we can determine
$\bf P$ and $\bf Q$ by solving Eq. (27).

Now we can determine the series of matrices $\textbf{B}_i$ and $\textbf{D}_i$
by solving the following equation
\begin{equation}
\left\{
\begin{aligned}
&{\bf B}_1+{\bf B}_2+{\bf B}_3+{\bf B}_4=2{\bf I},\\
&{\bf D}_1+{\bf D}_2+{\bf D}_3+{\bf D}_4=2{\bf I},\\
&{\bf B}_1-{\bf B}_3={\bf P},\\
&{\bf D}_1-{\bf D}_3=-{\bf P},\\
&{\bf B}_2-{\bf B}_4={\bf Q},\\
&{\bf D}_2-{\bf D}_4=-{\bf Q},\\
&{\bf B}_1+{\bf B}_3+{\bf D}_1+{\bf D}_3={\bf M}+{\bf P}^2,
\end{aligned} \right.
\end{equation}
where
\begin{equation}
{\bf M}=\left(
  \begin{array}{ll}
     c_1^2+c_2^2 &  \\
   & c_1^2\\
  \end{array}
\right).
\end{equation}

Need to emphasize again that $c_1$ and $c_2$ have the ability to describe
the elementary properties of elastomer. Therefore, the CA model is
able to describe the elementary properties of an elastic body.

Through selecting a series of $p_0$, $p_1$, $q_0$, and $q_1$ as follows
\begin{equation}\left \{
\begin{aligned}
 p_0 &=-q_1=\sqrt{1-\frac{1}{2} c_1^2},\\
 p_1 &=q_0=\sqrt{1-\frac{1}{2} c_1^2-\frac{1}{2} c_2^2}.
\end{aligned}\right .
\end{equation}
we can obtain a CA model with the following parameter matrices
\begin{equation}
\left\{
\begin{aligned}
{\bf B}_1={\bf D}_3&=\frac{1}{4}({\bf M}+2{\bf P}+{\bf P}^2),\\
{\bf B}_2={\bf D}_4&=\frac{1}{4}(4{\bf I}-{\bf M}-{\bf P}^2+2{\bf Q}),\\
{\bf B}_3={\bf D}_1&=\frac{1}{4}({\bf M}-2{\bf P}+{\bf P}^2),\\
{\bf B}_4={\bf D}_2&=\frac{1}{4}(4{\bf I}-{\bf M}-{\bf P}^2-2{\bf Q}).
\end{aligned} \right.
\end{equation}
Especially, when $c_1=\sqrt{2}$ and $c_2=0$, the matrices
$\mathbf{B}_i=\mathbf{D}_i=\frac{1}{2} {\bf I}$. Thus, the model degrades into Chopard's model.

\subsection{Stability analysis of the Cellular Automata model}

The numerical scheme of this CA model can be written as follows
\begin{equation}
\left\{
\begin{aligned}
{\bf b}^+ &=-{\bf b}+ \sum_i {\bf B}_i S_i {\bf w},\\
{\bf w}^+ &=-{\bf w}+ \sum_j {\bf D}_j S_j (-{\bf b}+ \sum_i {\bf B}_i S_i{\bf w}).
\end{aligned} \right.
\end{equation}
We change Eq. (32) into matrix form
\begin{equation}
\left (
\begin{array}{c}
{\bf b}^+\\
{\bf w}^+
\end{array}
\right )
= \left(
\begin{array}{ll}
-{\mathbf I}              &  \sum_i     {\mathbf B}_i S_i\\
\sum_j {\mathbf D}_j S_j  &  \sum_{i,j} {\mathbf D}_j S_j {\mathbf B}_i S_i-{\mathbf I}
\end{array}
\right)
\left (
\begin{array}{c}
{\bf b}\\
{\bf w}
\end{array}
\right ).
\end{equation}
Performing Fourier transform to Eq. (33), we have
\begin{equation}
\left (
\begin{array}{c}
{\bf b}_k^+\\
{\bf w}_k^+
\end{array}
\right )={\bf G}_k \left (
\begin{array}{c}
{\bf b}_k\\
{\bf w}_k
\end{array}
\right),
\end{equation}
where
\begin{equation}
{\bf G}_k=\left( \begin{array}{ll}
-{\bf I}                & \sum_i     {\bf B}_i S_{ki}\\
\sum_j {\bf D}_j S_{kj} & \sum_{i,j} {\bf D}_j S_{kj} {\bf B}_i S_{ki}-{\bf I}
\end{array} \right).
\end{equation}
Here ${\bf B}_j S_{kj}={\bf B}_j e^{i {\bf k} \cdot {\bf
\Delta}_j}$, ${\bf D}_j S_{kj}={\bf D}_j e^{i {\bf k} \cdot {\bf
\Delta}_j}$, ${\bf k}$ is the wave vector, and ${\bf \Delta}_j$ is
the displacement direction vector.

Selecting appropriate matrices ${\textbf{B}_i}$ and ${\textbf{D}_i}$, we
solve the eigenvalues of ${\bf G}_k$
\begin{equation}
\left | \begin{array}{ll}
-{\bf I}-\lambda {\mathbf I}         &\,\,\,\,\,\, \sum_i{\bf B}_i S_{ki}\\
\sum_j {\bf D}_j S_{kj}              &\,\,\,\,\,\, \sum_{i,j} {\bf D}_j S_{kj} {\bf B}_i S_{ki}-{\bf I}
-\lambda {\mathbf I} \end{array} \right |=0,
\end{equation}
and let $|\lambda _i|=1$, thus we can obtain a conservative linear
CA model. The model described by Eq. (3), Eq. (24) and Eqs. (29)-(31) is
a conservative linear system. Namely, this model presents the ideal
wave process in linear elastomer.

\section{Numerical Examples}
In this section, several numerical examples are used to validate the CA model. The former two describe the translational
motion in elastomer, and the latter one simulates the single-mode motion
in elastic body.

\subsection{Examples of translational motion}

\begin{figure}[tbp]
\center{\epsfig{file=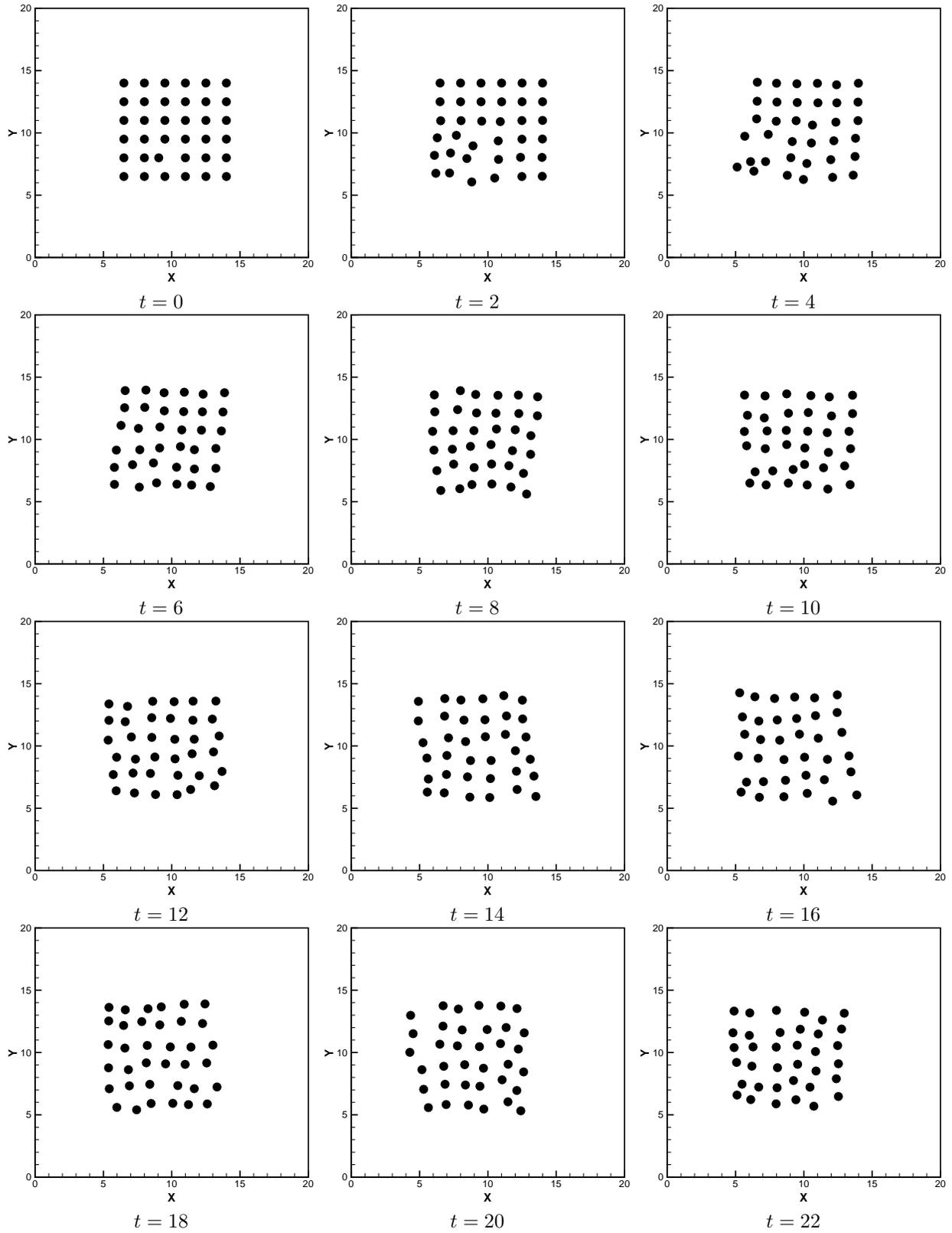,bbllx=106pt,bblly=95pt,bburx=577pt,bbury=712pt,
width=1.0\textwidth,clip=}}
\caption{CA simulations of the translational motion within the region.}
\end{figure}

\begin{figure}[tbp]
\center{\epsfig{file=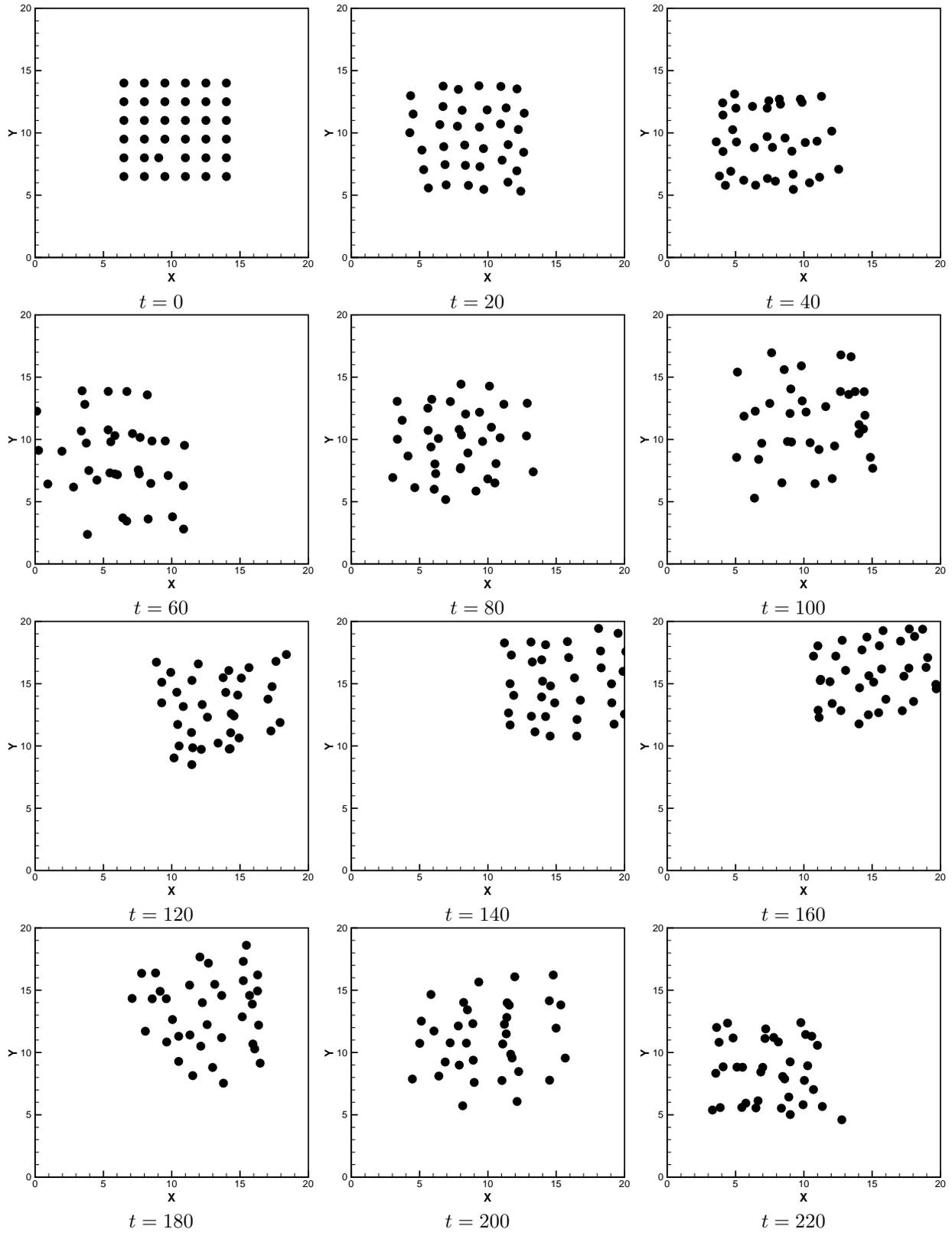,bbllx=106pt,bblly=95pt,bburx=577pt,bbury=712pt,
width=1.0\textwidth,clip=}}
\caption{CA simulations of the translational motion with bounce to boundaries.}
\end{figure}

In the former two examples, we select a square area $20\times20$ as the computing domain.
The size of mass is $6\times 6$. Here we set $c_1=0.4$, $c_2=0.58$.
Boundary conditions (BCs) are implemented as follows: (i) For the solid to bounce,
any $x$ or $y$ motion over the boundary is simply forbidden;
(ii) Free BCs are imposed on object which has some ghost particles on the boundary.

When the distance in $x$ or $y$ direction between the neighboring two particles equals
to a constant $\Delta_0$, the mass is static. Here $\Delta_0=1.5$. The initial
disturbance is in the following form: $x'(3,2)=x(3,2)-0.2\Delta_0$, as shown in Fig. 3.
It is noteworthy to mention that the initial disturbance is only added in $x$ direction.

There are two numerical examples about translational motion which share the same initial state.
Example (I) presents the evolution of the CA model within the region, as shown in Fig. 3.
Example (II) presents the translational motion with bounce to boundaries, as shown in Fig. 4.
In Example (I), although the disturbance is only added in $x$ direction, from Fig. 3 we can see that
particles of the mass present motions in $y$ direction.
The couple between different directions is illustrated more clearly in Example (II).
As a result of the contact with the boundary, not only the particles
have motions in $x$ direction, but the whole mass has motion in $y$ direction (see Fig. 4).
Therefore, we conclude that the CA model can couple the motion in $x$ direction with the motion in $y$ direction.

\subsection{Example of single-mode motion}

\begin{figure}[tbp]
\center{\epsfig{file=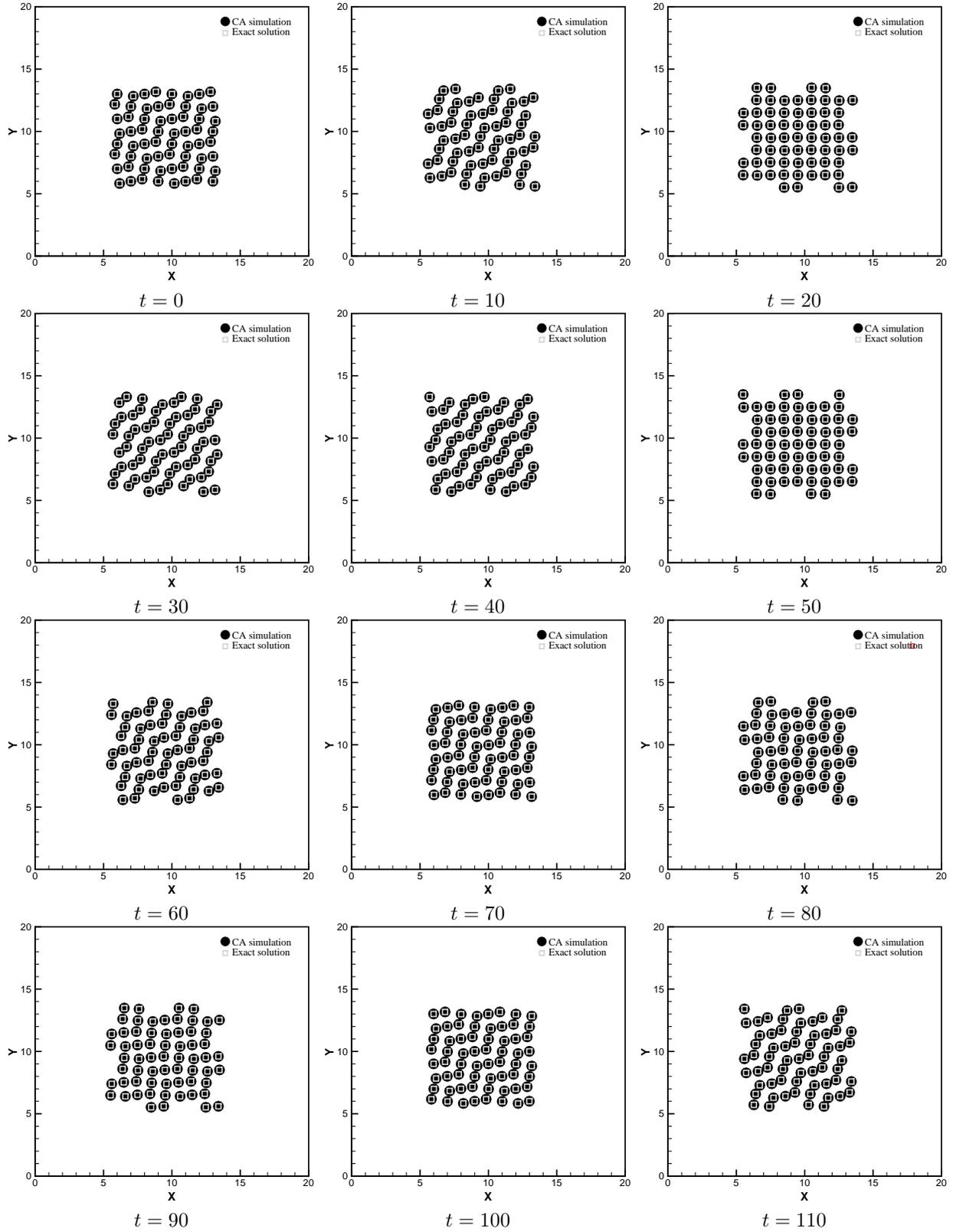,bbllx=106pt,bblly=95pt,bburx=577pt,bbury=712pt,
width=1.0\textwidth,clip=}}
\caption{Comparisons between CA simulations and the exact solutions.}
\end{figure}

To further validate the model, an example of single-mode motion
is illustrated here. In this example we select a square area
$20\times20$ as the computing domain. The size of object is $8\times 8$.
In this case, we set $c_1=0.5$ and $c_2=1.0$.

In this example, $\Delta_0 =1.0$ when the mass is static. To
obtain the single-mode motion, we select the eigenvalue $\lambda$ and the
corresponding eigenvectors ${\bf b}_{k0}$ and ${\bf w}_{k0}$ as follows
\begin{equation}
    \lambda=0.75+i0.661438,
    \end{equation}
    \begin{equation}
    \left (
    \begin{array}{c}
    a_{11}\\
    a_{12}\\
    a_{21}\\
    a_{22}
    \end{array}
    \right )= \left (
    \begin{array}{c}
    0.176777+i0.467707\\
    -0.176777-i0.467707\\
    -0.5\\
    0.5
    \end{array}
    \right ).
    \end{equation}

The initial displacements of the particles in mass are given by
the following equation
 \begin{equation}
    \left (
    \begin{array}{c}
    b_{0x}\\
    b_{0y}\\
    w_{0x}\\
    w_{0y}
    \end{array}
    \right )=Re \left( \left (
    \begin{array}{c}
    a_{11}\\
    a_{12}\\
    a_{21}\\
    a_{22}
    \end{array}
    \right ) e^{i{\bf k}\cdot {\bf r}}\right),
    \end{equation}
where $\bf k$ are the wave vectors, and $\bf r$ are the
situation vectors of every particle before the mass is disturbed.
From the view of analytic solution, the motion of the mass will be ruled by the following equation
\begin{equation}
    \left (
    \begin{array}{c}
    b_{x}(t)\\
    b_{y}(t)\\
    w_{x}(t)\\
    w_{y}(t)
    \end{array}
    \right )=Re \left( \left (
    \begin{array}{c}
    a_{11}\\
    a_{12}\\
    a_{21}\\
    a_{22}
    \end{array}
    \right ) e^{i ({\bf k}\cdot {\bf r}+\omega t)}\right),
    \end{equation}
where $\omega =-i \ln\lambda$.

The numerical scheme is ruled by Eq. (3), Eq. (24) and
Eqs. (29)-(31). The initial configuration is given by Eq. (39),
as shown in Fig. 5. The BCs are simpler than the above two examples.
Here we need not to consider the contact between mass and boundary of
the computing domain. BCs of the mass are periodic. Comparisons between CA solutions and the exact solutions are
illustrated in Fig. 5. It is clear that the two sets of results have a satisfying agreement.
Further, we draw the conclusion that the CA model has
the ability to present qualitatively wave propagation in elastomer.

It is necessary to point out that although the proposed model recovers the Possion ratio effects, compared with the deformation uniformity and periodicity of the true elastomer, the CA simulation results (see Figs. 3-5) still exist some gaps. These gaps firstly arise from the lack of introducing appropriate rules describing rotaion and elastic-plastic of solids in the CA model. Secondly, it should be emphasized that what the CA model concerns are different with other macroscopic numerical methods (e.g., finite-element method). The CA approach focuses on understanding of the physical mechanism behind the macroscopic physical phenomena, but not the comparisons with specific experimental data. Indeed, the CA approach can be regarded as coarse and large time step sampling results of the elastic body during its movement. Therefore, there are some differences between the CA simulation results and the realistic elastomer.

Meanwhile, we should note that the present work has already pointed out the way how to introduce the above-mentioned physical characteristics into the CA model. In other words, the new CA model can be further modified to describe more realistic dynamical behaviors in solid materials. For example, via relating the coefficient matrices to the displacement field, the CA model can be used to describe the nonlinear elastic phenomena; via changing the coefficients of  $\mathbf{b}$ and $\mathbf{w}$ from the fixed value, $-1$, the CA model can be used to describe the inner dissipation in the plastic behavior of the solid body; via breaking the neighboring relations of particles, the CA model can be used to describe damage and fracture phenomena.
Generally speaking, since the mesoscopic collision rules of the CA model and the macroscopic behaviors of solid exhibit complex nonlinear relationships, looking for mesoscopic rules which can be used to describe macroscopic behaviors (rotation, elastic-plastic, etc.) of solid is still an open problem. Recently, the progress has been rather slow and there is rare report on this issue. So, from this point of view, this paper provides us with an effective way to propose more powerful CA models and study dynamics of solids via this more essential and fundamental approach. This is another contribution of the present work.

\section{Conclusions}

A new two-dimensional cellular automaton model for solid dynamics is proposed. The solid body is represented by a set of white and black particles alternatively positioned in the $x$- and $y$- directions. The force acting on each particle is represented by the linear summation of relative displacements of the nearest-neighboring particles.  The key technique in this new model is the construction of eight coefficient matrices.
Theoretical and numerical analyses show that the present model can be mathematically described by a conservative system. So, it is a model for elastic material. In the continuum limit the CA model recovers the well-known Navier equations. The components of the coefficient matrices can be related to the elastic parameters of material, such as the Lame coefficients or the shear module and Poisson ratio.

Chopard \emph{et al.} introduced the CA model into the field of solid dynamics for the first time. Even though in the form of two dimensions, the original CA model by Chopard \emph{et al.} is in fact quasi-one dimensional and can only describe some simple wave phenomena.
The new model presented in this paper realizes the natural coupling of deformations in the $x$- and $y$- directions. Consequently, the wave phenomena related to the Poisson ratio effects are successfully recovered. When the eight coefficient matrices are all diagonal, the new model goes back to the original CA model by Chopard \emph{et al.}.
This work advances significantly the CA modeling and simulation in the field of computational solid dynamics.

\section*{ACKNOWLEDGEMENTS}
The authors would like to sincerely thank Drs. Bo Yan, Feng Chen, Weiwei Pang, Wei Li and Chuandong Lin for helpful discussions.



\end{document}